\documentstyle[12pt]{article}

\begin{document}
\setcounter{page}{1}
\pagestyle{plain} \vspace{1cm}
\begin{center}
\Large{\bf  Coherent States of Harmonic Oscillator and Generalized Uncertainty Principle}\\
\small
\vspace{1cm}
{\bf Kourosh Nozari}\quad and \quad {\bf Tahereh Azizi }\\
\vspace{0.5cm}
{\it Department of Physics,
Faculty of Basic Science,\\
University of Mazandaran,\\
P. O. Box 47416-1467,
Babolsar, IRAN\\
e-mail: knozari@umz.ac.ir}
\end{center}
\vspace{1.5cm}
\begin{abstract}
In this paper dynamics and quantum mechanical coherent states of a
simple harmonic oscillator are considered in the framework of
Generalized Uncertainty Principle(GUP). Equations of motion for
simple harmonic oscillator are derived and some of their new
implications are discussed. Then coherent states of harmonic
oscillator in the case of GUP are compared with relative situation
in ordinary quantum mechanics. It is shown that in the framework of
GUP there is no considerable difference in definition of coherent
states relative to ordinary quantum mechanics. But, considering
expectation values and variances of some operators, based on quantum
gravitational arguments one concludes that although it is possible
to have complete coherency and vanishing broadening in usual quantum
mechanics, gravitational induced uncertainty destroys complete
coherency in quantum gravity and it is not possible to have a
monochromatic ray in principle. \\
{\bf PACS}: 04.60.-m, 11.17.+y\\
{\bf Keywords}: Quantum Gravity, Generalized Uncertainty Principle,
Quantum Dynamics, Harmonic Oscillator and Coherent States
\end{abstract}
\newpage

\section{Introduction}
Recently it has been indicated that measurements in quantum
gravity are governed by generalized uncertainty principle. There
are some evidences from string theory and black holes Physics,
based on gedanken experiments, which leads some authors to
re-examine usual uncertainty principle of Heisenberg[1]. These
evidences have origin on the quantum fluctuation of the background
spacetime metric. Introducing of this idea has drown attention and
many authors considered various problems in the framework of
generalized uncertainty principle[2-14]. Such investigations have
revealed that actually spacetime is not commutative, gravity is
not fundamental interaction, constants of the nature are not
constant and notion of locality and position space representation
are not satisfied in Planck scale. Therefore, it seems that a
re-formulation of quantum theory should be
performed to incorporate gravitational effects in planck scale.\\
The problem of harmonic oscillator in the context of GUP first has
been considered by Kempf {\it et al}[15]. They have found
eigenvalues and eigenfunctions of harmonic oscillator in the
context of GUP by direct solving of the Schr$\ddot{o}$dinger
equation. Then Camacho has analyzed the role that GUP can play in
the quantization of electromagnetic field. He has considered
electromagnetic
oscillation modes as simple harmonic oscillations[16].\\
In this paper we first consider the problem of dynamics for harmonic
oscillator in the framework of GUP. Then quantum mechanical coherent
states of a simple harmonic oscillator are examined in GUP. Some
arguments, based on quantum gravitational viewpoint, will show that
although there is no difference between the representation and
definition  of  coherent states in the GUP and usual quantum
mechanics, quantum gravitational uncertainty will destroy the notion
of complete coherency. This will be shown by calculating expectation
values of momentum and position operators and their variances.
Note that this is a novel implication of GUP since in usual
quantum mechanics it is possible to have complete coherency in principle.\\
The structure of the paper is as follow: section 2 provides an
overview to GUP. Section 3 is devoted to dynamics of simple
harmonic oscillator in GUP. Section 4 considers quantum mechanical
coherent states of harmonic oscillator. The paper follows by
summary in section 5.
\section{A Generalized Uncertainty Principle } Usual uncertainty
principle of quantum mechanics, the so-called Heisenberg
uncertainty principle, should be re-formulated because of the
non-commutative nature of space-time. This statement leads to
existence of a minimal observable distance on the order of Plank
length $l_{P}$ in quantum gravity. In the context of string
theories, this observable distance is referred to GUP. A
generalized uncertainty principle can be formulated as
\begin{equation}
\label{math:2.1}\Delta x\geq\frac{\hbar}{\Delta p}+const.G\Delta
p,
\end{equation}
which, using the minimal nature of $ l_{P} $ can be written as,
\begin{equation}
\label{math:2.2} \Delta x\geq\frac{\hbar}{\Delta p}+l_{p}^2
\frac{\Delta p}{\hbar}.
\end{equation}
The corresponding Heisenberg commutator now becomes,
\begin{equation}
\label{math:2.3} [x,p]=i\hbar(1+\beta p^2)
\end{equation}
Actually as Kempf {\it et al} have argued[15], one can consider
more generalization such as
\begin{equation}
\label{math:2.4} \Delta x\Delta
p\geq\frac{\hbar}{2}\Big(1+\alpha(\Delta x)^2+\beta(\Delta
p)^2+\gamma \Big)
\end{equation}
and the corresponding commutator relation is
\begin{equation}
\label{math:2.5}[x,p]=i\hbar(1+\alpha x^2+\beta p^2).
\end{equation}
The main consequence of these GUPs is that measurement of position
is possible only up to Plank length, $ l_{P} $. So one can not
setup a measurement to find more accurate particle position than
Plank length.

\section{Dynamics of Harmonic Oscillator in GUP}
For simplicity we consider GUP as equation (3). In Heisenberg
picture of quantum mechanics, equation of motion for observable
$A$ is as follow,
\begin{equation}
\label{math:3.1}\frac{dA}{dt}= \frac{i}{\hbar}[H,A].
\end{equation}
Hamiltonian for a simple harmonic oscillator is,
\begin{equation}
\label{math:3.2}H=\frac{p^2}{2m}+ \frac{1}{2}m\omega^{2} x^2
\end{equation}
Now the equations of motion for $x$ and $p$ are respectively,
\begin{equation}
\label{math:3.3}\frac{dx}{dt}= \frac{1}{m}\Big(p+\beta p^3\Big),
\end{equation}
and
\begin{equation}
\label{math:3.4}\frac{dp}{dt}=
-\frac{1}{2}m\omega^{2}\Big(2x+\beta x p^2 +\beta p^2 x\Big).
\end{equation}
Using Baker-Hausdorff lemma, a lengthy calculation gives the
following equations for time evolution of $x$ and $p$
respectively,

$$x(t) = x(0)\cos\omega t+ \frac{p(0)}{m\omega}
\sin \omega t$$
$$+ \beta \Bigg[\frac{p^{3}(0)}{m\omega}(\omega t) -
\frac{1}{2}\Bigg(p(0)x(0)p(0)+\frac{3}{2}\Big[x(0)p^{2}(0)+
p^{2}(0)x(0)\Big] \Bigg)(\omega t)^2$$
 $$- \Bigg(\frac{5}{6}
\frac{p^{3}(0)}{m\omega}-\frac{5}{12} m\omega\Big[x^{2}(0)p(0)+
p(0)x^{2}(0)\Big] - \frac{1}{2}m\omega x(0)p(0)x(0)\Bigg)(\omega
t)^3$$
\begin{equation}
\label{math:3.5} + \Bigg(\frac{11}{24} \Big[x(0)p^{2}(0) +
p^{2}(0) x(0)\Big] +
\frac{5}{12}p(0)x(0)p(0)-\frac{1}{3}m^{2}\omega^{2} x^{3}(0)\Bigg)
(\omega t )^4 \Bigg],
\end{equation}
and
$$p(t) = p(0)\cos\omega t- m\omega x(0)\sin \omega t$$
$$+ \beta \Bigg[-\frac{1}{2}m\omega \Big[x(0)p^{2}(0)+
p^{2}(0)x(0)\Big](\omega t)$$
$$-\Bigg( p^{3}(0) -\frac{1}{4}
m^{2}\omega^{2}\Big[ p(0)x^{2}(0)+ x^{2}(0)p(0)
+2x(0)p(0)x(0)\Big]\Bigg)(\omega t)^2$$
\begin{equation}
\label{math:3.5} +\Bigg(\frac{2}{3} m\omega\Big[x(0)p^{2}(0)+
p^{2}(0)x(0)\Big]+ \frac{1}{2}p(0)x(0)p(0)-\frac{1}{3}
m^{3}\omega^{3} x^{3}(0)\Bigg)(\omega t)^{3}\Bigg],
\end{equation}
where only terms proportional to first order of $\beta$ are
considered. It is evident that in the limit of $\beta\rightarrow
0$ one recover the usual results of ordinary quantum mechanics.
The term proportional to $\beta$ shows that in the framework of
GUP harmonic oscillator is no longer "harmonic" since, now its
time evolution is not oscillatory completely. Now for computing
expectation values, we need a well-defined physical state. Note
that eigenstates of position operators are not physical states
because of existence of a minimal length which completely destroys
the notion of locality. So we should consider a physical state
such as $|\alpha>$ where $|\alpha>$ is for example a momentum
space eigenstate[15]. Suppose that
$p_{\alpha}(0)=<\alpha|p(0)|\alpha>$ and
$x_{\alpha}(0)=<\alpha|x(0)|\alpha>$. Now the expectation value of
momentum operator is,
$$\frac{<\alpha|p(t)|\alpha>}{m} = \frac{p_{\alpha}(0)}{m}\cos\omega t- \omega x_{\alpha}(0)\sin \omega t$$
$$+ \beta \Bigg[-\frac{1}{2}\omega \Big(x_{\alpha}(0)p_{\alpha}^{2}(0)+
p_{\alpha}^{2}(0)x_{\alpha}(0)\Big)(\omega t)$$
$$-\Bigg( \frac{p_{\alpha}^{3}(0)}{m} -\frac{1}{4}
m\omega^{2}\Big[ p_{\alpha}(0)x_{\alpha}^{2}(0)+
x_{\alpha}^{2}(0)p_{\alpha}(0)
+2x_{\alpha}(0)p_{\alpha}(0)x_{\alpha}(0)\Big]\Bigg)(\omega t)^2$$
\begin{equation}
\label{math:3.6} +\Bigg(\frac{2}{3}
\omega\Big[x_{\alpha}(0)p_{\alpha}^{2}(0)+
p_{\alpha}^{2}(0)x_{\alpha}(0)\Big]+
\frac{1}{2m}p_{\alpha}(0)x_{\alpha}(0)p_{\alpha}(0)-\frac{1}{3}
m^{2}\omega^{3} x_{\alpha}^{3}(0)\Bigg)(\omega t)^{3}\Bigg].
\end{equation}
This relation shows that there is a complicated dependence of the
expectation value of momentum operator to the mass of the
oscillator. In usual quantum mechanics,
$\frac{<\alpha|p(t)|\alpha>}{m}$ and $ \frac{p_{\alpha}(0)}{m}$ are
mass independent. Here although $ \frac{p_{\alpha}(0)}{m}$ is still
mass independent, but now $\frac{<\alpha|p(t)|\alpha>}{m}$ has a
complicated mass dependence. This is a novel implication which have
been induced by GUP. Physically, it is completely reasonable that
the expectation value for momentum of a particle be a function of
its mass, but the mass dependence here has a complicated form
relative to usual situation.
\section{Coherent States of Harmonic Oscillator}
We consider the simple harmonic oscillator by hamiltonian
\begin{equation}
\label{math:2.6} H=\frac{1}{2m}(p^2+m^2\omega^2x^2)
\end{equation}
The problem of quantum oscillator is easily solved in terms of the
annihilation and creation operators $a$ and $a^{\dagger}$. We
recall the fundamental definitions:
\begin{equation}
\label{math:2.7}
a=\sqrt{\frac{m\omega}{2\hbar}}\big(x+\frac{ip}{m\omega}\big),
\end{equation}
\begin{equation}
\label{math:2.7}a^{\dagger}=\sqrt{\frac{m\omega}{2\hbar}}\big(x-\frac{ip}{m\omega}\big)
\end{equation}
and the inverse relations:
\begin{equation}
\label{math:2.8}x=\sqrt{\frac{\hbar}{2m\omega}}(a+a^{\dagger}),\qquad
p=i\sqrt{\frac{m\hbar\omega}{2}} (-a+a^{\dagger}).
\end{equation}
The Hamiltonian H is given in terms of these operators as :
\begin{equation}
\label{math:2.9}H=\hbar\omega(a^{\dagger}a+\frac{1}{2})
\end{equation}
If we set $ N\equiv a^{\dagger} a $ (: Number operator),then
\begin{equation}
\label{math:2.10}[N,a^{\dagger}]=a^{\dagger}, \quad
[N,a]=-a,\qquad [a^{\dagger},a]=-1
\end{equation}
Let $\textbf{H}$ be a Fock space generated by $a$ and $
a^{\dagger}$, and $\{|n\rangle| n\in \textbf\{N\}\cup\{0\}\}$ be
its basis. The action of $a$ and $a^{\dagger}$ on $\textbf{H}$ are
given by
\begin{equation}
\label{math:2.11}a|n\rangle=\sqrt{n}|n-1\rangle, \qquad
a^{\dagger}|n\rangle=\sqrt{n+1}|n+1\rangle, \qquad
N|n\rangle=n|n\rangle
\end{equation}
Where $|0\rangle$ is a normalized vacuum ($a|0\rangle = 0$  and
$\langle0|0\rangle = 1$). Therefore states $|n\rangle$ for $ n\geq1$
are given by
\begin{equation}
|n\rangle=\frac{{a^{\dagger}}^n}{\sqrt{n!}}|0\rangle.
\end{equation}
These states satisfy the orthogonality and completeness conditions
\begin{equation}
\label{math:2.12}\langle m|n\rangle =\delta_{mn}, \qquad
\sum_{n=0}^{\infty}|n\rangle\langle n|=1.
\end{equation}
By definition, coherent state is the normalized state
$|\lambda\rangle \in \textbf{H}$, which is the eigenstate of
annihilation operator and satisfies the following equation,
\begin{equation}
 \label{math:2.13}a|\lambda\rangle =\lambda|\lambda\rangle \qquad where \qquad \langle\lambda|\lambda\rangle=1
\end{equation}
and
\begin{equation}
\label{math:2.13}|\lambda\rangle=e^{-|\lambda|^2/2}\sum_{n=0}^{\infty}\frac{\lambda^n}
{\sqrt{n!}}|n\rangle=e^{-|\lambda|^2/2}e^{\lambda
a^{\dagger}}|0\rangle.
\end{equation}
Actually $\lambda$ can be complex because $a$ is not Hermittian.
The coherent state was introduced by Schr$\ddot{o}$dinger as the
quantum state of the harmonic oscillator which minimizes the
uncertainty equally distributed in both position $x$ and momentum
$p$. Let us now consider the following possibility, as a
generalization for these two operators in GUP,
\begin{equation}
\label{math:3.1}  a=\frac{1}{\sqrt{2\hbar\omega}}\bigg(\omega
x+i[p+f(p)]\bigg),
\end{equation}
\begin{equation}
\label{math:3.2}
a^{\dagger}=\frac{1}{\sqrt{2\hbar\omega}}\bigg(\omega
x-i[p+f(p)]\bigg).
\end{equation}
Here $f(p)$ is a function that satisfies three conditions, namely:
(i) in the limit $\beta\rightarrow0 $ we recover the usual
definition for the creation and annihilation operators,(14) and
(15); (ii) if $\beta\neq0$, then we have (3), and; (iii)
$[a_{\vec{k}},a^{\dagger}_{\vec{k^{\prime}}}]=i\hbar
\delta_{\vec{k}\vec{k^{\prime}}}$. It can be shown that the
following function satisfies the aforementioned restriction
\begin{equation}
\label{math:3.3}
f(p_{\vec{k}})=\sum_{n=1}^{\infty}\frac{(-\beta)^n}{2n+1}p_{\vec{k}}^{2n+1}
\end{equation}
Condition (iii) means that the usual results, in relation with the
structure of the Fock space, are valid in our case, for instance,
the definition of the occupation number operator, $N_{\vec{k}}=
a^{\dagger}_{\vec{k}} a_{\vec{k}}$ , the interpretation of
$a^{\dagger}_{\vec{k}}$ and $a_{\vec{k}}$ are creation and
annihilation operators, respectively, etc. Clearly, the relation
between $p_{\vec{k}}$, $a_{\vec{k}}$ and $a^{\dagger}_{\vec{k}}$ is
not linear, and from the Hamiltonian (13) we now deduce that it is
not diagonal in the occupation number representation. Let us now
consider
\begin{equation}
\label{math:3.4} f(p_{\vec{k}})=-\frac{\beta}{3}p_{\vec{k}}^3
\end{equation}
In this form we find $p_{\vec{k}}$ as a function of $a_{\vec{k}}$
and $a^{\dagger}_{\vec{k}}$, namely
\begin{equation}
\label{math:3.5}
p_{\vec{k}}=-i\sqrt{\frac{\hbar\omega}{2}}\big(a_{\vec{k}}-a^{\dagger}_{\vec{k}}\big)
\big[1-\sqrt{\frac{\hbar\omega\beta}{8}}(a_{\vec{k}}-a^{\dagger}_{\vec{k}})\big]
\end{equation}
It is clear that, if $\beta=0$ we recover the usual case.
Rephrasing the Hamiltonian as a function of the creation and
annihilation operators we find:
\begin{equation}
\label{math:3.6}
H=\sum_{\vec{k}}\hbar\omega\big[N_{\vec{k}}+\sqrt{\frac{\hbar\omega\beta}{8}}
g(a_{\vec{k}},a^{\dagger}_{\vec{k}})+\beta\frac{(\hbar\omega)^2}{16}h(a_{\vec{k}},a^{\dagger}_{\vec{k}})\big]
\end{equation}
where functions $ g(a_{\vec{k}},a^{\dagger}_{\vec{k}})$ and
$h(a_{\vec{k}},a^{\dagger}_{\vec{k}}) $ are:
\begin{equation}
\label{math:3.7}g(a_{\vec{k}},a^{\dagger}_{\vec{k}})=a_{\vec{k}}^3-N_{\vec{k}}a_{\vec{k}}
-a_{\vec{k}}N_{\vec{k}}-a_{\vec{k}}-(a^{\dagger}_{\vec{k}})^3+N_{\vec{k}}
a^{\dagger}_{\vec{k}}+a^{\dagger}_{\vec{k}}N_{\vec{k}}+a^{\dagger}_{\vec{k}}
\end{equation}
and
\begin{equation}
\label{math:3.8}
    h(a_{\vec{k}},a^{\dagger}_{\vec{k}})=a_{\vec{k}}^4+
 a_{\vec{k}}^2(a^{\dagger}_{\vec{k}})^2-a_{\vec{k}}^3a^{\dagger}_{\vec{k}}-
 a_{\vec{k}}^2a^{\dagger}_{\vec{k}}a_{\vec{k}} $$ $$ +(a^{\dagger}_{\vec{k}})^2a_{\vec{k}}^2+
 (a^{\dagger}_{\vec{k}})^4-
(a^{\dagger}_{\vec{k}})^2a_{\vec{k}}a^{\dagger}_{\vec{k}}-
(a^{\dagger}_{\vec{k}})^3a_{\vec{k}}$$
$$ -a_{\vec{k}}a^{\dagger}_{\vec{k}}a_{\vec{k}}^2-
a_{\vec{k}}(a^{\dagger}_{\vec{k}})^3+a_{\vec{k}}a^{\dagger}_{\vec{k}}a_{\vec{k}}a^{\dagger}_{\vec{k}}
+a_{\vec{k}}(a^{\dagger}_{\vec{k}})^2a_{\vec{k}} $$
$$ -a^{\dagger}_{\vec{k}}a_{\vec{k}}^3-a^{\dagger}_{\vec{k}}a_{\vec{k}}(a^{\dagger}_{\vec{k}})^2+
a^{\dagger}_{\vec{k}}a_{\vec{k}}^2a^{\dagger}_{\vec{k}}+
a^{\dagger}_{\vec{k}}a_{\vec{k}}a^{\dagger}_{\vec{k}}a_{\vec{k}}.
\end{equation}
Now with these pre-requisites we can consider the coherent states
in the context of GUP. Suppose $|\lambda\rangle$ be an eigenstate
of the annihilation operator. We remember that the definition of
the annihilation operator in GUP may be different from the usual
quantum mechanics but the fact that eigenstates of annihilation
operator are coherent states do not changes. Therefore one can
write
\begin{equation}
\label{math:3.9}   a|\lambda\rangle=\lambda|\lambda\rangle
\end{equation}
Indeed $|n\rangle$ is the eigenstate of the number operator and
satisfies completeness and orthogonality conditions. So we can
expand $|\lambda>$ in terms of the stationary states $|n\rangle$
\begin{equation}
\label{math:3.10}
|\lambda\rangle=\sum_{n=0}^{\infty}|n\rangle\langle
n|\lambda\rangle=C_{n}|n\rangle,
\end{equation}
The eigenvalue equation(19) implies the following recursion
formula for the expansion coefficients:
\begin{equation}
\label{math:3.11}C_{n}=\frac{\lambda}{\sqrt{n}}C_{n-1},
\end{equation}
We immediately obtain
\begin{equation}
\label{math:3.12}C_{n}=\frac{\lambda^{n}}{\sqrt{n!}} C_{0},
\end{equation}
The constant $ C_{0}$ is determined from the normalization
condition on the Fock space,
\begin{equation}
\label{math:3.13}1=\langle\lambda|\lambda\rangle
=|C_{0}|^2\sum_{n=0}^{\infty}\frac{\lambda^{2n}}{\sqrt{n!}}=|C_{0}|^2e^{|\lambda|^2},
\end{equation}
For any complex number $\lambda$ the correctly normalized
quasi-classical state $|\lambda\rangle$ is therefore given by
\begin{equation}
\label{math:3.14}|\lambda\rangle=e^{-\frac{1}{2}|\lambda|^2}\sum\frac{|\lambda|^n}{\sqrt{n!}}|n\rangle,
\end{equation}
We recall that the n-th stationary state$|n\rangle$ is obtained
from the ground state wave function by repeated application of the
operator $a^{\dagger}$,
\begin{equation}
\label{math:3.15}|n\rangle=\frac{1}{\sqrt{n!}}(a^{\dagger})^n|0\rangle,
\end{equation}
This allows us to write the coherent state in the form:
\begin{equation}
\label{math:3.16}|\lambda\rangle=e^{-\frac{1}{2}|\lambda|^2}\sum_{n=0}^{\infty}\frac{1}{n!}(\lambda
a^{\dagger})^n|0\rangle = e^{-\frac{1}{2}|\lambda|^2} e^{\lambda
a^{\dagger}}|0\rangle
\end{equation}
We see that this expression for the eigenstates of the annihilation
operator is the same as usual quantum mechanics, equation (23).
Actually, it is not surprising that there is no changes in the form
of states by modifying the uncertainty relation and similarly for
the coherent state. The unchanged state itself cannot be the result
of considering generalized uncertainty principle (GUP). It is
because a quantum state does not necessarily imply a direct
connection with uncertainty principle. Differences caused by
different uncertainty relations (such as the GUP) will be found in
the expectation values of the operators for a given state and their
statistics(such as variance) that can be obtained from the
measurement on the state. To analyze the coherent state under the
GUP, we should consider $<x>$ and $<p>$ for the coherent state and
see that whether they are changed or not. For this end, suppose that
$|\lambda>$ is a coherent state given by (39). Since
\begin{equation}
\label{math:5.1}x=\sqrt{\frac{\hbar}{2\omega}}(a_{\vec{k}}
+a^{\dagger}_{\vec{k}}),
\end{equation}
and
\begin{equation}
\label{math:5.2}p=-i\sqrt{\frac{\hbar\omega}{2}}(a_{\vec{k}}-a^{\dagger}_{\vec{k}})
[1-\sqrt{\frac{\hbar\omega\beta}{8}}(a_{\vec{k}}-a^{\dagger}_{\vec{k}})],
\end{equation}
one finds the following result for the expectation value of position
operator, $x$
\begin{equation}
\label{math:5.3}<x>=<\lambda|x|\lambda>=\sqrt{\frac{\hbar}{2\omega}}
<\lambda|a_{\vec{k}}+a^{\dagger}_{\vec{k}}|\lambda>=
\sqrt{\frac{\hbar}{2\omega}}(\lambda+\lambda^{\ast}).
\end{equation}
Therefore one has,
\begin{equation}
\label{math:5.4}<x>^2=\frac{\hbar}{2\omega}(\lambda^2+{\lambda^{\ast}}^2
+2\lambda\lambda^{\ast})=\frac{\hbar}{2\omega}(\lambda+\lambda^{\ast})^2.
\end{equation}
It is straightforward to show that,
\begin{equation}
\label{math:5.5}<x^2>={\frac{\hbar}{2\omega}}(\lambda^2+{\lambda^{\ast}}^2+
2\lambda\lambda^{\ast}+1)=\frac{\hbar}{2\omega}(\lambda+\lambda^{\ast})^2+1,
\end{equation}
and therefore we find for the variance of $x$,
\begin{equation}
\label{math:5.6}(\Delta x)^2=<x^2>-<x>^2=\frac{\hbar}{2\omega}.
\end{equation}
This is the same as usual quantum mechanics result. This is not
surprising since the definition of position operator is the same as
its definition in usual quantum mechanics.\\
In the same manner, a simple calculation gives,
\begin{equation}
\label{math:5.7}<p>=-i\sqrt{\frac{\hbar\omega}{2}}\Big[(\lambda-\lambda^{\ast})
-\sqrt{\frac{\hbar\omega\beta}{8}}[(\lambda-\lambda^{\ast})^2-1]\Big],
\end{equation}
and
$$<p>^2=-\frac{\hbar\omega}{2}\Big\{(\lambda-\lambda^{\ast})^2
-2\sqrt{\frac{\hbar\omega\beta}{8}}(\lambda-\lambda^{\ast})[(\lambda-
\lambda^{\ast})^2-1]+$$
\begin{equation}
\label{math:5.8}
\frac{\hbar\omega\beta}{8}[(\lambda-\lambda^{\ast})^2-1]^2\Big\}.
\end{equation}
Since,
\begin{equation}
\label{math:5.9}p^2=-\frac{\hbar\omega}{2}\Big[(a_{\vec{k}}-a^{\dagger}_{\vec{k}})^2
-2\sqrt{\frac{\hbar\omega\beta}{8}}(a_{\vec{k}}-a^{\dagger}_{\vec{k}})^3+\frac{\hbar
\omega\beta}{8}(a_{\vec{k}}-a^{\dagger}_{\vec{k}})^4\Big],
\end{equation}
then,
\begin{equation}
\label{math:5.10}
<p^2>=-\frac{\hbar\omega}{2}\Big\{[(\lambda-\lambda^{\ast})^2-1]-$$
$$2\sqrt{\frac{\hbar\omega\beta}{8}}(\lambda^3-{\lambda^{\ast}}^3-
3\lambda^{\ast}\lambda^2+3{\lambda^{\ast}}^2\lambda+3\lambda^{\ast}-3\lambda)+$$
$$\frac{\hbar\omega\beta}{8}(\lambda^4+{\lambda^{\ast}}^4-4\lambda^{\ast}\lambda^3-
4{\lambda^{\ast}}^3\lambda+6{\lambda^{\ast}}^2\lambda^2-6\lambda^2-6{\lambda^{\ast}}^2+
12\lambda^{\ast}\lambda+3)\Big\},
\end{equation}
and therefore one finds,
\begin{equation}
\label{math:5.11}(\Delta
p)^2=<p^2>-<p>^2=-\frac{\hbar\omega}{2}[-1-2\sqrt{\frac{\hbar\omega\beta}{8}}
(2\lambda^{\ast}-2\lambda)+\frac{\hbar\omega\beta}{8}(-4\lambda^2-4{\lambda^{\ast}}^2
+8\lambda^{\ast}\lambda+2)],
\end{equation}
or by some manipulations, one obtains the following result for
variance of $p$,
\begin{equation}
\label{math:5.12}({\Delta
p})^2=\frac{\hbar\omega}{2}+\hbar\omega\sqrt{\frac{\hbar\omega\beta}{2}}
(\lambda^{\ast}-\lambda)+\frac{\hbar^2\omega^2\beta}{8}[1-2(\lambda^{\ast}-\lambda)^2]
\end{equation}
Note that these results give the usual quantum mechanical results
when $\beta \longrightarrow 0$. Equations (46) and (51) show that
although the definition of coherent states do not changes in GUP,
but because of quantum gravitational effect one can not have
coherent state in principle. Now there is a considerable departure
from very notion of coherency. In usual quantum mechanics one can
have complete coherency in principle. One can localize wave pocket
in space completely, at least in principle and wave can propagate
without broadening in principle. This is evident from $\Delta
x\geq\frac{\hbar}{\Delta p}$. In quantum gravity because of
gravitational induced uncertainty, one can not localize wave pocket
at all and it is impossible to omit broadening. Therefore in quantum
gravity one can not have any solitonic states and any wave pocket
will be broad at least by gravitational effects.

\section{Summery}
In this paper, we have obtained the equation of motion for a simple
harmonic oscillator in the framework of GUP using Heisenberg picture
of quantum mechanics. In fact in this situation the terminology of
"simple" is no longer applicable because of nonlinear terms in
equations of dynamics. It has been shown that there is a complicated
mass dependence of momentum operator expectation value which has
considerable departure from usual quantum mechanical result. This is
a novel implication of GUP. Also by a simple calculation we have
shown that the definition of the coherent states is not different in
the framework of GUP with usual quantum mechanics. But there is some
considerable difference due to gravitational uncertainty. Because of
these uncertainties, in quantum gravity one can not have solitonic
states and any wave pocket will be broad  by gravitational effects.
The main result of our calculations, specially computation of
expectation values and variances, is that in the framework of GUP
very notion of coherency breaks down.\\
{\bf Acknowledgement}\\
We would like to appreciate an unknown referee for him valuable
comments and careful reading of the manuscript. Actually we are
indebted to him in some sense.

\end{document}